# Un nouvel algorithme pour la détection des transferts horizontaux de gènes partiels entre les espèces et pour la classification des transferts inférés

Alix Boc*, Alpha Boubacar Diallo* et Vladimir Makarenkov*

*Département d'informatique
Université du Québec à Montréal
Case postale 8888, Succursale Centre-ville
Montréal (Québec) H3C 3P8 Canada
makarenkov.vladimir@uqam.ca

**Résumé.** Dans cet article, nous présentons un nouvel algorithme pour la détection et la validation des transferts horizontaux de gènes (THG) partiels. L'algorithme proposé se base sur une procédure de fenêtre coulissante analysant les fragments d'un alignement de séquences multiples. Une procédure de validation par bootstrap, permettant d'évaluer le support statistique des THG partiels obtenus, a été introduite. Le nouvel algorithme peut être utilisé pour confirmer ou rejeter les transferts complets détectés en utilisant n'importe quel algorithme de détection des THG et ainsi pour classifier les transferts retrouvés (en tant que complets ou partiels). Il peut aussi être appliqué à une échelle génomique pour estimer la proportion de gènes mosaïques dans chaque génome étudié, de même que le taux d'échange génétique entre les espèces impliquées.

## 1 Introduction

Les bactéries et les archées s'adaptent à différentes conditions environnementales via la formation de gènes mosaïques. Le terme "mosaïque" découle de la configuration des blocs entrecoupés de séquences ayant des histoires évolutionnaires différentes, mais se trouvant combinées dans le gène résultant, à la suite d'évènements de recombinaison intragénique (Zhaxybayeva *et al.*, 2004). Les segments recombinés peuvent être dérivés d'autres souches de la même espèce ou des espèces différentes (Gogarten *et al.*, 2002).

Le transfert horizontal de gènes, qui est un des principaux mécanismes de l'évolution réticulée (Makarenkov et Legendre 2000; Makarenkov *et al.*, 2004), est responsable de l'apparition de gènes mosaïques. Le modèle du transfert horizontal complet suppose que soit le gène transféré supplante le gène orthologue entier dans le génome receveur, soit, si le gène transféré est absent du génome receveur, il lui est ajouté (Boc *et al.*, 2010). Le second modèle, celui du transfert partiel, implique la formation des gènes mosaïques. Un gène mosaïque est formé à travers les mécanismes de la transformation et de la conjugaison qui permettent l'acquisition et l'intégration subséquente de fragments d'ADN provenant des organismes distincts. Alors que plusieurs méthodes ont été proposées pour l'identification et la validation des THG complets (Page 1994; Mirkin *et al.*, 1995 ; Maddison, 1997; Lawrence et Ochman, 1997; Hallett et Lagergren, 2001 ; Boc et Makarenkov 2003 ; Tsirigos et Rigoutsos 2005; Than et Nakhleh 2008 et Boc *et al.*, 2010), seulement deux algorithmes traitent le problème de l'inférence de THG partiels (Denamur *et al.*, 2000 et Makarenkov *et al.*, 2006).

Toutefois, les deux derniers travaux ne considèrent pas la problématique de la validation des transferts partiels obtenus et n'incluent pas de simulations Monte-Carlo. Dans les faits, aucune méthode fiable pour l'*identification des gènes mosaïques* et des *transferts horizontaux de gènes partiels* associés n'a été proposée jusqu'à maintenant.

## 2  Algorithme pour la détection des transferts partiels

Les principales étapes de l'algorithme, qui cherche à produire un scénario optimal (i.e., comportant le nombre minimal de transferts) de transferts partiels d'un gène donné pour un groupe d'espèces considérées, sont résumées ci-dessous. Une validation par bootstrap est effectuée pour chaque transfert partiel obtenu et seuls les transferts significatifs (i.e., ayant le pourcentage de bootstrap supérieurs à un seuil donné) sont inclus dans la solution finale. Une procédure de fenêtre coulissante est utilisée pour tester différents fragments de l'alignement de séquences multiples (ASM). Notons qu'une approche utilisant une fenêtre coulissante a été précédemment utilisée pour détecter les recombinaisons (Ray 1998; Lee et Sung, 2008), mais aucun de ces travaux ne traite le problème de transferts partiels.

*Pas préliminaire.* Soit $X$ un ensemble d'espèces étudiées, ASM un alignement de séquences multiples de taille $l$ et $S_{i,j}$ le fragment de l'ASM, étant analysé, situé entre les sites $i$ et $j$, où $1 <= i < j <= l$. Définissons aussi la taille de la fenêtre coulissante $w$ ($w = j - i+1$) et la taille du pas de progression $s$. Inférons l'*arbre phylogénétique d'espèces T*. D'habitude, un arbre basé sur les caractères morphologiques ou sur une molécule qui est supposée être réfractaire aux transferts horizontaux de gènes (e.g., 16S sRNA) joue le rôle de l'arbre d'espèces. L'arbre $T$ doit être enraciné en respectant les hypothèses d'évolution connues. Fixons la taille de la fenêtre coulissante $w$ et la taille du pas $s$ (dans nos simulations, les tailles de fenêtres égales à $l/5$, $l/4$, $l/3$, $l/2$ et le pas de progression de 10 sites ont été utilisés).

*Pas k.* Fixons la position de la fenêtre coulissante dans l'intervalle $[i,j]$, où $i = 1 + s(k - 1)$ et $j = i + w - 1$. Si $i + w - 1 > l$, alors $j = l$. Inférons un *arbre de gène partiel T'* caractérisant l'évolution du fragment de l'ASM localisé dans l'intervalle $[i,j]$. Dans cette étude, la méthode *PhyML* (Guindon et Gascuel, 2003) a été employée pour inférer les arbres de gène partiels. Appliquons un algorithme de détection existant pour inférer un scénario de THG partiels associés à l'intervalle $[i, j]$. Ici, nous avons utilisé l'algorithme *HGT-Detection* décrit dans Boc *et al.* (2010) pour inférer des transferts complets, mais n'importe quel autre algorithme pourrait être utilisé à sa place. Cet algorithme de détection des THG est plus rapide et dans la plupart des cas aussi précis que les populaires algorithmes *LatTrans* (Hallett and Lagergren, 2001) et *RIATA-HGT* (Than et Nakhleh 2008). La dissimilarité de bipartition (Boc *et al.* 2010) ou la distance de Robinson et Foulds (Makarenkov et Leclerc 1996 ; Makarenkov et Leclerc 2000) peuvent être utilisées à cette étape en tant que critères d'optimisation de *HGT-Detection*. Une procédure servant à évaluer la fiabilité des transferts partiels obtenus a aussi été développée. Cette procédure, basée sur le principe de bootstrap, prend en compte l'incertitude des arbres de gène partiels, ainsi que le nombre de fois qu'un transfert donné apparait dans tous les scénarios de coût minimal (i.e., comportant le nombre minimum de THG nécessaires pour réconcilier les arbres $T$ et $T'$).

*Pas final.* Établissons une liste des transferts partiels prédits. Identifions les *intervalles entrelacés* donnant lieu à des transferts partiels identiques (i.e., les mêmes donneurs et receveurs et la même direction). Ré-exécutons l'algorithme de détection des THG pour tous les intervalles entrelacés (considérant leur longueur totale dans chaque cas) produisant les *THG* *partiels identiques*. Si ces THG partiels sont trouvés à nouveau (c'était habituellement le cas dans nos simulations) pour un fragment de séquences plus long englobant les intervalles entrelacés, évaluons leur support de bootstrap et, dépendamment du support obtenu, incluons les ou non dans la solution finale. La complexité de cet algorithme est comme suit :

$$O(r \times (\frac{(l-w)}{s} \times (C(PhIn) + \tau \times n^4))), \quad (1)$$

où $w$ est la taille de la fenêtre coulissante, $s$ est le pas de progression, $C(PhIn)$ est la complexité de la méthode d'inférence d'arbres phylogénétiques (e.g., *PhyML*) utilisée pour inférer les phylogénies à partir des fragments de séquences situés dans la fenêtre coulissante, $r$ est le nombre de réplicats dans le bootstrap, $n$ est le nombre d'espèces et $\tau$ est le nombre moyen de transferts horizontaux détectés pour un fragment de séquences de taille $w$.

*Simulations Monte-Carlo*. Des simulations Monte-Carlo ont été effectuées pour tester l'efficacité du nouvel algorithme dans le contexte des THG partiels. Nous avons examiné comment l'algorithme proposé se comporte en fonction du nombre d'espèces observées. La procédure de simulations incluait les étapes suivantes. Des arbres d'espèces binaires avec 8, 16, 32 et 64 feuilles ont été créés en utilisant la procédure de génération d'arbres aléatoires de Kuhner et Felsenstein (1994). Les longueurs des arêtes des arbres ont été générées en utilisant une distribution exponentielle. Nous avons ensuite exécuté le programme *SeqGen* (Rambaut et Grassly, 1996) pour générer des alignements de séquences multiples de protéines le long des arêtes des arbres d'espèces construits à la première étape. Des séquences de protéines avec 500 acides aminés ont été générées. Puis, pour chaque arbre d'espèces $T$, nous avons généré des arbres de gènes $T'$, ayant le même nombre de feuilles, en effectuant des déplacements aléatoires de ses sous-arbres. Un modèle satisfaisant toutes les contraintes d'évolution plausibles a été implémenté pour générer les THG partiels aléatoires. Pour chaque arbre d'espèces, 1 à 5 déplacements aléatoires de sous-arbres ont été effectués et différents arbres de gènes $T'$, englobant entre 1 et 5 THG partiels, ont été générés. Pour chaque arbre de gène, les fragments de séquences dans les sous-arbres affectés par les THG ont été régénérés avec *SeqGen*. Nous avons fixé la taille de chaque séquence transférée à 200 acides aminés. Les alignements de séquences multiples obtenus contenaient donc des blocs de séquences affectées par des THG. Finalement, nous avons exécuté l'algorithme pour chaque arbre d'espèces généré et l'alignement de séquences multiples associé qui était affecté par les THG partiels. La taille de la fenêtre coulissante a été fixée à 100, 200, 300, 400, puis 500 acides aminés ; 100 réplicats de chaque arbre de gène partiel $T'$ ont été générés pour évaluer le support de bootstrap des arêtes de $T'$ dans un premier temps, puis le support des THG partiels obtenus dans un deuxième temps. Les arbres qui avaient le support de bootstrap inférieur à 60% ont été retirés de l'analyse. Parmi les THG obtenus, seuls les transferts avec un bootstrap supérieur à 90% ont été retenus. Finalement, nous avons estimé le taux de détection (les vrais positifs seulement) et le taux de faux positifs.

Les performances moyennes obtenues par le nouvel algorithme sont illustrées sur la figure 1. Pour chaque ensemble de paramètres (taille de l'arbre, nombre de THG générés), 100 jeux de données répliqués ont été testés. La figure 1 montre que les meilleurs taux de détection ont été obtenus pour les arbres de 16 espèces. Cette figure met en lumière les différences entre le taux de détection moyen et le taux de faux positifs moyen en fonction du nombre d'espèces. La moyenne ici était calculée à partir des résultats obtenus pour les arbres englobant 1 à 5 THG générés. Alors que le taux de détection moyen était toujours supérieur à 70% (79,6% en moyenne), le taux moyen de faux positifs était toujours inférieur à 40% (30,8% en moyenne). Ces résultats suggèrent que le nouvel algorithme peut être utile pour détecter des THG partiels et ainsi pour identifier des gènes mosaïques.

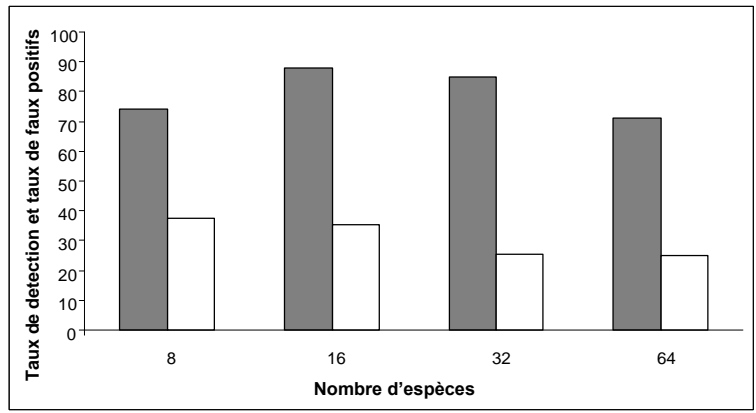

FIG. 1 – *Taux moyens de détection (vrais – colonnes grises et faux positifs – colonnes blanches) des transferts partiels générés.*

## 3 Conclusion

Nous avons décrit un nouvel algorithme pour la prédiction des transferts horizontaux de gènes partiels et pour l'identification des gènes mosaïques. Au meilleur de notre connaissance, ce problème d'actualité n'a pas été convenablement traité dans la littérature. L'algorithme proposé se base sur une procédure de fenêtre coulissante pour analyser les fragments de l'alignement de séquences multiples. La taille de la fenêtre coulissante doit être ajustée en fonction des informations existantes sur les gènes et les espèces étudiés. Une procédure de validation, permettant d'évaluer la robustesse des transferts partiels obtenus et prenant en compte l'incertitude des arbres de gène partiels, a aussi été développée. Le nouvel algorithme peut être utilisé pour confirmer ou exclure les transferts complets inférés avec n'importe quel algorithme de détection des THG et ainsi pour classifier les transferts retrouvés (en tant que complets ou partiels). Cet algorithme peut aussi être appliqué à une échelle génomique pour estimer la proportion de gènes mosaïques dans des génomes des espèces étudiées, de même que pour déterminer le taux de transferts complets et partiels entre ces espèces. Plusieurs statistiques intéressantes concernant la position et la fonctionnalité des fragments génétiques affectés par les transferts horizontaux, aussi bien que le taux de transferts inter- et intra-espèces, peuvent être calculées en utilisant la technique discutée. L'algorithme proposé a été inclus dans le package *T-REX* (Makarenkov, 2001) disponible sur : www.trex.uqam.ca.

Algorithme pour la détection des transferts horizontaux de gènes partiels

**Résumé en anglais**. In this article we are describing a new algorithm for detecting and validating partial horizontal gene transfers (HGT). The presented algorithm is based on a sliding window procedure which analyzes fragments of the given multiple sequence alignment. A bootstrap procedure incorporated in our method can be used to estimate the support of each inferred partial HGT. The new algorithm can be also applied to confirm or discard complete (i.e., traditional) horizontal gene transfers detected by any HGT inferring algorithm. While working on a full-genome scale, the introduced algorithm can be used to assess the level of mosaicism of the whole species genomes as well as the rates of complete and partial HGT underlying the evolution of the considered set of species.